\documentclass[final]{svjour3}
\usepackage{graphicx}
\usepackage{rotating}
\usepackage{amssymb}
\usepackage{mathptmx}
\usepackage[numbers]{natbib}
\makeatletter
\journalname{Journal of Low Temperature Physics}

\begin{document}

\newcommand{\hdblarrow}{H\makebox[0.9ex][l]{$\downdownarrows$}-}
\title{Operation of an archaeological lead PbWO$_4$ crystal to search for neutrinos from astrophysical sources with a Transition Edge Sensor}

\author{$^{1,2}$N.~Ferreiro Iachellini \and $^{3,4}$L.~Pattavina \and $^{1}$A.~H.~Abdelhameed \and $^{1,5}$A.~ Bento \and $^{1}$L.~Canonica \and $^{6}$F.~Danevich \and $^{7}$O.~M.~Dubovik \and D.~Fuchs$^{1}$ \and $^{1}$A.~Garai \and $^{1}$M.~Mancuso \and $^{1}$F.~Petricca \and $^{7}$I.~A.~Tupitsyna}
\institute{$^1$Max-Planck-Institut für Physik, M\"unchen, Germany D-80336\\
$^2$Exzellenzcluster ORIGINS, Garching, Germany, D-85748\\
$^3$INFN Laboratori Nazionali del Gran Sasso, Assergi, Italy, 67100\\
$^4$Physik-Department, Technische Universität München, Garching, Germany, 85748\\
$^5$LIBPhys, Departamento de Fisica, Universidade de Coimbra, Coimbra, Portugal, P-3004 516\\
$^6$Institute for Nuclear Research of NASU, Kyiv, Ukraine, 02000\\
$^7$Institute for Scintillation Materials, Kharkiv, Ukraine, 61072}

\maketitle

\begin{abstract}

The experimental detection of the CE$\nu$NS allows the investigation of neutrinos and neutrino sources with all-flavor sensitivity. Given its large content in neutrons and stability, Pb is a very appealing choice as target element. The presence of the radioisotope $^{210}$Pb (T$_{1/2}\sim$22 yrs) makes natural Pb unsuitable for low-background, low-energy event searches. This limitation can be overcome employing Pb of archaeological origin, where several half-lives of $^{210}$Pb have gone by. We present results of a cryogenic measurement of a 15g PbWO$_4$ crystal, grown with archaeological Pb (older than $\sim$2000 yrs) that achieved a sub-keV nuclear recoil detection threshold. A ton-scale experiment employing such material, with a detection threshold for nuclear recoils of just 1 keV would probe the entire Milky Way for SuperNovae, with equal sensitivity for all neutrino flavors, allowing the study of the core of such exceptional events.

\keywords{Cryogenic particle detector, Supernovae neutrinos, Energy Resolution, Coherent Neutrino Nucleus Scattering}

\end{abstract}

\section{Introduction}

Core-collapse supernovae (CC-SNe) are one of the most powerful phenomena occurring in our Universe. These dramatic transients may mark the death of stars heavier than $8\ M_\odot$ and result in the formation of neutron stars and black holes\cite{janka:2006fh}. SNe are a unique laboratory to investigate the properties of particles and matter under extreme conditions, the formation of chemical elements and the birth of new stars. The understanding of the physics driving the Core-Collapse is therefore of great interest for particle physics and astrophysics\cite{Janka:2016fox}. The role of neutrinos in driving these events is of particular importance as they are key to trigger the process and are messengers of the nuclear reactions that take place during the explosion.

The recent discovery of coherent elastic neutrino-nucleus scattering\cite{Akimov:2017ade} (CE$\nu$NS) offers a new intriguing detection channel for SNe neutrinos. Driven by a Z-boson exchange, CE$\nu$NS is a neutral current process and, as such, offers equal sensitivity to all neutrino flavors. It is a low energy scattering where an impinging neutrino with energy of $O$(10MeV) transfers $O$(1keV) momentum to the recoiling nucleus\cite{Drukier:1983gj}. The appeal of this process as detection channel comes from the fact that at such low energy the momentum transfer is of the same order of the de Broglie wavelength of the target nucleus and, therefore, the cross-section gets coherently enhanced. This coherence enhancement results in a cross-section that is orders of magnitude larger than the ones of Inverse Beta Decay (IBD) and Electron Scattering (ES), allowing relatively small detectors to assess SNe emission properties with the same precision as the one of gigantic neutrino telescopes, provided that they achieve a detection threshold capable of assessing these low energy recoils\cite{PhysRevD.102.063001}.

\section{Detection of SNe neutrinos}
\subsection{The coherent neutrino nucleus scattering}\label{sec:cnns}
The cross-section for CE$\nu$NS can be derived within the Stanard Model and it reads:
\begin{equation}
\label{eq:xsec}
\frac{d\sigma}{d E_R} = \frac{G^2_F m_N}{8 \pi (\hslash c )^4} \left[(4\sin^2 \theta_W -1 ) Z + N\right]^2 \left(2- \frac{E_R m_N}{E^2}\right) \cdot |F(q)|^2,\ 
\end{equation}
$G_F$ is the Fermi coupling constant, $\theta_W$ the Weinberg angle, $Z$ and $N$ the atomic and neutron numbers of the target nucleus, $m_N$ its mass, $E$ the  energy of the incoming neutrino and $E_R$ the recoil energy of the target. The term $F(q)$ represent the distribution of the weak chanrge within the nucleus at momentum transfer $q=\sqrt{2E_R m_N}$.
From Eq.~(\ref{eq:xsec}) we observe that large $N$ and $m_N$ favor the scattering. Heavy elements, such as Pb, are therefore the most interesting ones to be used as target.

The average recoil energy $E_R$ can be derived from Eq.~(\ref{eq:xsec}):
\begin{equation}
\label{eq:eravg}
\langle E_R \rangle = \frac{2E^2}{3m_N},\
\end{equation}
We can estimate the average recoiling energy knowing that SNe neutrinos have energies $O$(10MeV), so that $E_R\sim 1$keV.

The detection of $O$(1keV) nuclear recoils is an experimental challenge that has been tackled by direct dark matter searches in the last decades and low-background cryogenic particle detectors are the most promising technology due to their excellent energy resolution that, thanks to the calorimetric nature of their operation, directly translates to energy detection threshold well below 1keV\cite{CRESST:2015txj}.
\subsection{Cryogenic detection of SNe neutrinos}
In the following, we investigate scintillating PbWO$_4$ crystals as target for SNe neutrinos because of the presence of Pb that leverages the coherent enhancement of the CE$\nu$NS cross-section and its crystalline form that offers good performance to the operation as cryogenic detector. 

The total number of events induced in a detector with energy detection threshold $E_{th}$ and number of target nuclei $N_T$ can be written as:
\begin{eqnarray}
N_{\rm{exp}} &=&  \int dt\int_{E_{\rm{th}}} \frac{d^2N}{dE_R\; dt} dE_R =\\ \nonumber
&=&\sum_{i} N_{\rm{T}} \int dt \int_{E_{\rm{th}}} f_i(E,t) \; \frac{d\sigma}{dE_R}  \;dE\ , \label{eq:rate}
\end{eqnarray}
where $f_i(E,t)$ is the neutrino flux on Earth for the $ith$ flavor at the time $t$.
For a CC-SN of $27\ M_\odot$ occurring at 10~kpc from Earth for 1.7~ton of PbWO$_4$ $N_{exp}$ is 16.4events per ton of target material. Such detector would have the capability to probe the entire Milky Way for SNe of that type\cite{RES-NOVA:2021gqp}. 

The radioisotope $^{210}$Pb naturally present in Pb has an activity of several Bq/kg, making the usage of natural Pb unfit for the purpose of SNe neutrino detection. In order to overcome this issue, we study the employment of Pb of archaeological origin as target material, where several $T_{1/2}$ of $^{210}$Pb have gone by and the residual activity of this dangerous background is $\le$715~$\mu$Bq/kg, resulting in a count rate of 0.1~counts/keV/ton/10s for recoil energies in the range 1-40keV\cite{PhysRevD.102.063001}.

The operation of  PbWO$_4$ crystal as scintillating cryogenic particle detector offers the possibility of further background suppression by the simulaneous measurement of scintillation light and deposited energy\cite{Beeman:2012wz}. The utilization of Transition Edge Sensors (TES) guarantees elevated sensitivity.

\section{Operation of an archaeological-Pb containing PbWO$_4$ crystal with a Transition Edge Sensor}
\subsection{Experimental set-up}
A scintillating PbWO$_4$ crystal grown utilizing Pb of archaeological origin of mass 15.7g has been equipped with a W Transition Edge Sensor (TES) of the type utilized to instrument the light detectors of the CRESST experiment\cite{Rothe:2018bnc}. The deposition of the W-film was done by magnetron sputtering\cite{abdelhameed2020deposition}. The crystal was produced by double crystallization from deeply purified archaeological lead and high purity tungsten oxide by Czochralski method at the Institute for Scintillation Materials of NASU. No dedicated sensor design optimization was performed.

The crystal is supported by a copper holder but not directly in contact with it. It is pressed against four sapphire balls with a metallic clamp, so that the contact surface of the crystal with the thermal bath is minimized. On the surface of the holder facing the crystal a $^{55}$Fe radioactive source was glued. 

\begin{figure}[htbp]
\begin{center}
\includegraphics[width=0.9\linewidth, keepaspectratio]{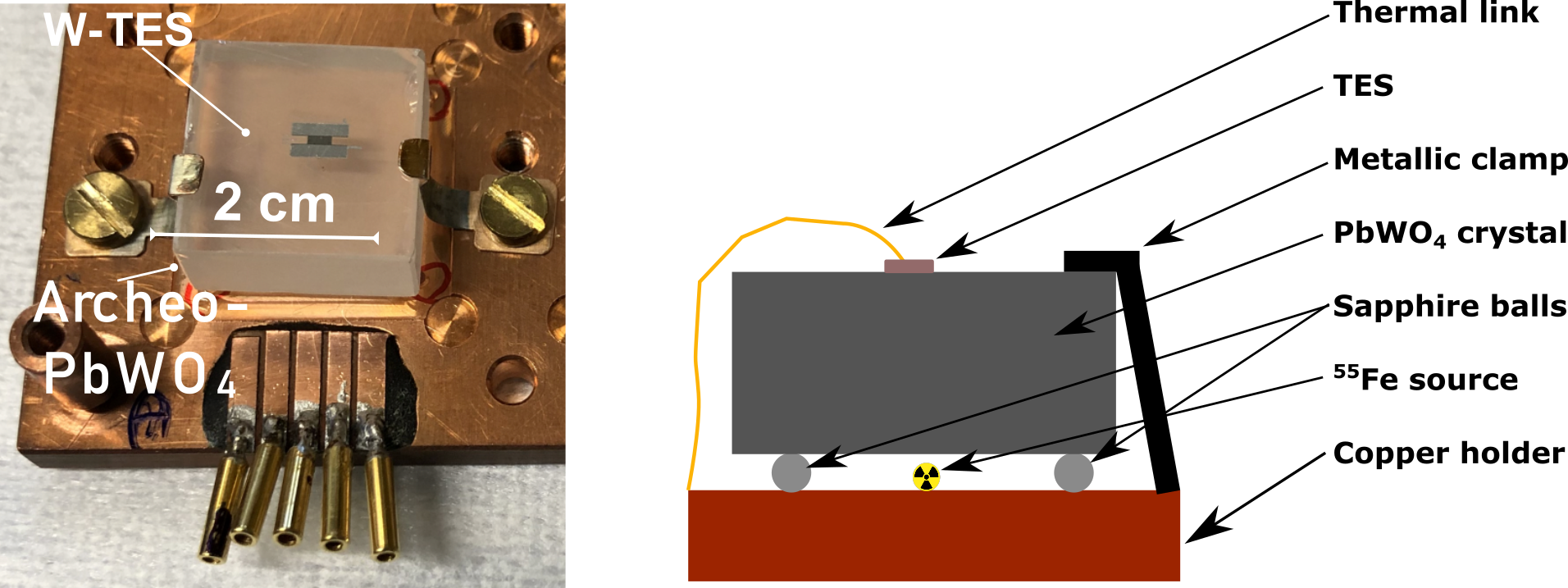}
\caption{(color online). Picture of the detector ({\it Left}) and schematics of the experimental set-up ({\it Right}). The detector is supported by a block of copper that provides the thermal coupling to the cryostat. The crystal is pressed against four small sapphire balls to reduce the surface contact to the copper holder. The TES is weakly coupled to the copper holder via thin gold wire. In between the crystal and the copper holder a $^{55}$Fe radioactive source is present.}
\end{center}
\label{fig:setup}
\end{figure}

The detector was operated for 70~hours in a dilution refrigerator at Max-Planck-Institut f\"ur Physik in Munich (see Fig.~\ref{fig:setup}). 

During operation, the TES was biased with a constant dc current of 1$\mu$A and read-out by a dc-squid. Its output was digitized and written to disk. During the whole operation of the detector, the TES operating point was constantly monitored and the detector's temperature adjusted accordingly, so that the TES response was kept as constant as possible during the measurement.

Pulses resulting from thermal excitations as well as randomly sampled empty baselines have been acquired. These latter are used to determine the noise power spectrum in the later analysis. The datasets generated and analysed during the current study are available from the corresponding author on reasonable request
\subsection{Energy calibration}
\begin{figure}[htbp]
\begin{center}
\includegraphics[width=0.9\linewidth, keepaspectratio]{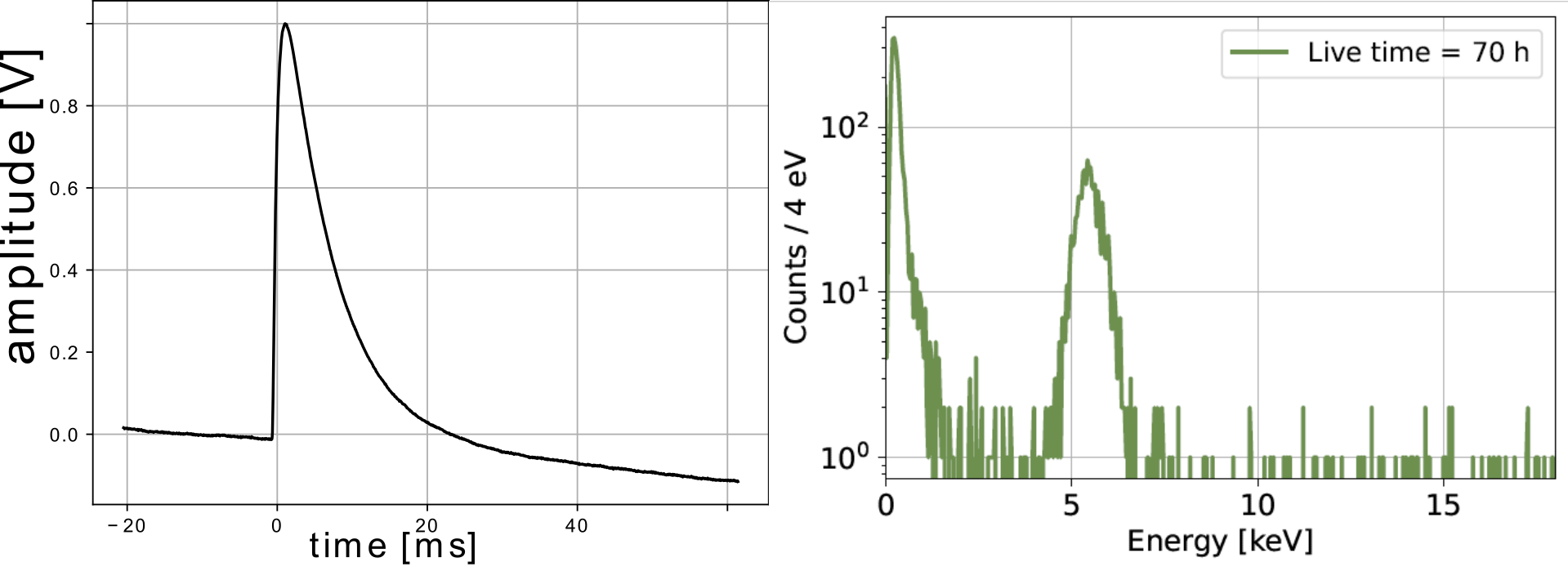}
\caption{(color online). ({\it Left}) Average pulse in operating point obtained from a list of selected pulses in the linear region of the sensor. Due to the elevated count rate during the measurement, the average pulse shows a tilted baseline, indicating that the baseline is not flat on the time-scale of a pulse. ({\it Right}) Calibrated energy spectrum recorded over a raw time of 70~hours. The prominent peak at 5.895keV is from the K$_\alpha$ gamma line of $^{55}$Mn and used to calibrate the energy spectrum.}
\end{center}
\label{fig:cal}
\end{figure}

The TES response to particle energy depositions was characterized averaging a number of pulses recorded from the detector. These pulses were selected to have an amplitude small enough to fall in the linear response range of the sensor. In addition, stability and quality cuts were applied to reject puile-up event, artifacts and unstable preriods. The resulting average pulse represents the detector's response in the linear region, free of read-out noise (Fig.~\ref{fig:cal} {\it Left}). The amplitude of each event is determined fitting the average pulse to its waveform. In the fitting procedure, waveform samples above the linear region of the TES are not considered. This counteracts the degrading of the fit due to the saturation of the TES, effectively extending the dynamic range of the detector.

In the amplitude spectrum the K$_\alpha$ line of the $^{55}$Mn is identified and its nominal value (5.895keV) used to calibrate the events (Fig.~\ref{fig:cal} {\it Right}). 
\subsection{Resolution and threshold}
\begin{figure}[htbp]
\begin{center}
\includegraphics[width=0.9\linewidth, keepaspectratio]{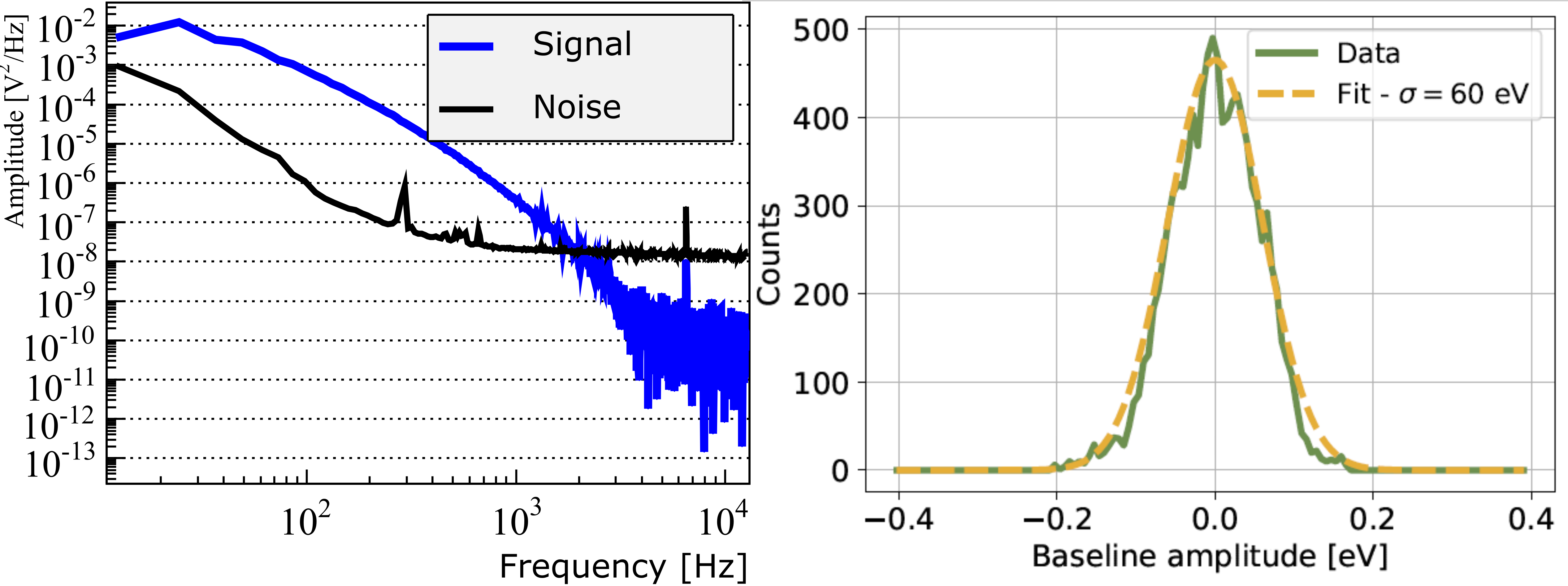}
\caption{(color online). {\it Left} Noise power spectrum of the average signal ({\it Blue} and of the noise ({\it Black}). These two quantities are used to compute the Optimum Filter that offers the optimal weighting of the signal frequency-components to achieve the best S/N ratio. {(\it Right}) Distribution of the empty baselines samples at the output of the filter.}
\end{center}
\label{fig:resolution}
\end{figure}

The baseline resolution of the detector was calculated using the Optimum Filter\cite{Gatti1986ProcessingTS}. The transfer function of the filter is calculated from the average pulse power spectrum and the noise power spectrum (Fig.\ref{fig:resolution} {\it Left}). The signal components get re-weighted with their Signal-to-Noise (R/N) ratio at each frequency, so that the filter's output has the highest S/N ratio and,thus, the optimal resolution. The waveform samples of the empty baseline at the output of the filter are gaussian distributed and we extract the baseline noise by fitting this distribution (Fig.\ref{fig:resolution} {\it Right}). We obtain a resolution of $\sigma$ = (60.3$\pm$4.1)eV. 

The detection threshold for a cryogenic particle detector can be set at any value, as long as the number of triggers due to random noise oscillation remains negligible compared to the sought-for signal. We conservatively estimate the energy threshold as $5\cdot\sigma$, obtaining E$_{th}$ = (301.5$\pm$20.5)eV, well below the 1keV value discussed in Section~\ref{sec:cnns}.

\section{Conclusions and outlook}
The next Galactic Supernova will be a precious occasion to shed light on the physics powering SNe and a high statistics, time-resolved detection of this event is paramount. We showed that PbWO$_4$ crystals operated as cryogenic detectors are an appealing option for the detection of SNe neutrinos and we operated a crystal of 15g, grown with archaeological lead, as cryogenic particle detector, proving that a PbWO$_4$ grown from archaeological material can be operated with the required energy threshold.

In order to fully characterize the potential of cryogenic PbWO$_4$ detectors, dedicated underground measurements are still necessary to assess the background level. In addition, an optimization of the TES design is required in order to operate detectors with a larger active mass.
\begin{acknowledgements}
This research was partially supported by the Excellence Cluster ORIGINS which is funded by the Deutsche Forschungsgemeinschaft (DFG, German Research Foundation) under Germany’s Excellence Strategy - EXC-2094 - 390783311.
\end{acknowledgements}

\pagebreak
\bibliographystyle{unsrt}
\bibliography{refs}

\end{document}